\documentclass[twocolumn]{aa}
\usepackage{graphicx}
\usepackage{txfonts}
\begin{document}
   \title{An empirical calibration of sulphur abundance in ionised gaseous nebulae}

   \author{E. P\'erez-Montero
          \inst{1}, A.I. D\'\i az
     \inst{1},
          J.M. V\'\i lchez\inst{2}, C. Kehrig
      \inst{2,3}
          }

   \offprints{E. P\'erez-Montero}

   \institute{Departamento de F\'\i sica Te\'orica, 
         C-XI, Universidad Aut\'onoma de Madrid,
              28049, Cantoblanco, Madrid, Spain.\\
              \email{enrique.perez@uam.es}\\
              \email{angeles.diaz@uam.es}
         \and
             Instituto de Astrof\'\i sica de Andaluc\'\i a (CSIC)
        Apartado de Correos 3004. 18080,  Granada, Spain.\\
             \email{jvm@iaa.es}\\  
              \email{kehrig@iaa.es}
	\and
             Observat\'{o}rio Nacional,
             Rua Jos\'{e} Cristino, 77, 20.921-400, Rio de Janeiro - RJ, Brazil\\
             \email{kehrig@on.br}
        }

   \date{}

   \abstract{
We have derived an empirical calibration of the abundance of S/H as a function of the
S$_{23}$ parameter, defined using the bright sulphur lines of [SII] and [SIII].
Contrary to what is the case for the widely used O$_{23}$ parameter, the calibration
remains single valued up to the abundance values observed in the disk HII regions.
The calibration is based on a large sample of nebulae for which direct determinations
of electron temperatures exist and the sulphur chemical abundances can be directly
derived. ICFs, as derived from the [SIV] 10.52 $\mu$ emission line (ISO observations),
are shown to be well reproduced by Barker's formula for a value of $\alpha$ = 2.5.
At any rate, only about 30 \% of the objects in the sample require ICFs larger than 1.2.
The use of the proposed calibration opens the possibility of performing abundance
analysis with red to IR spectroscopic data using S/H as a metallicity tracer.

   }
 
\authorrunning{P\'erez-Montero et al.}
   \maketitle
%

\section{Introduction}


Oxygen is the main abundance tracer in HII regions and HII galaxies, but its abundance is 
rather uncertain in those cases in which no direct determinations of the electron 
gas temperature exist therefore requiring the use of empirical or semi-empirical methods.
 These methods are based on the cooling properties
of ionised nebulae which ultimately reflect on a relation between
emission line intensities and oxygen abundance. In fact, when the
cooling is dominated by oxygen, the electron temperature depends
inversely on oxygen abundance. Since the intensities of collisionally
excited lines depend exponentially on temperature, a relation is
expected to exist between these intensities and oxygen abundances. The
O$_{23}$ parameter, also known as R$_{23}$ and defined as the sum of the intensities of the [OII]
$\lambda\lambda$ 3727,29 \AA\ and [OIII]  $\lambda\lambda$ 4959, 5007 \AA\ 
emission lines relative to
H$\beta$ (Pagel et al., 1979), has been widely used for these purposes. 

The relation between O$_{23}$ and oxygen abundance is however
two-folded 
since at high metallicities, the efficiency of the oxygen as a cooling agent
decreases the strength of the oxygen emission lines while at low 
metallicities the cooling is mainly exerted by hydrogen 
and the oxygen line strengths increase with metallicity. A value of
12+log(O/H) of about 8.2 divides the two different abundance regimes. Although some line 
ratios have been proposed in order to break this degeneracy, the fact that 
a large number of HII regions/HII galaxies lie right 
on the turnover region is of great concern.

On the other hand, the use of S/H as an abundance tracer has been
frequently overlooked.  Similarly to oxygen, sulphur is an element
produced in massive stars through explosive nucleosynthesis and its
yield should follow that of O. Nebular S/H abundances are therefore expected
to follow O/H and the S/O is expected to remain constant at 
approximately the solar neighborhood value, log S/O $\simeq$ -1.6
(e.g. Lodders 2003; Bresolin et al. 2004). Empirical tests exploring
this have been performed confirming the cosmic nucleosynthetic ratio
though the results of some works suggest that this relation should be
explored further, particularly at the not well known metallicity ends:
extremely metal deficient HII galaxies (i.e. very low O/H) and HII
regions in the inner disk of galaxies (i.e. metal rich central parts
with highest O/H abundances).

%
   \begin{table*}
      \caption[]{Reddening corrected line fluxes for sulphur normalized to I(H$\beta$)=100
measured in spectra of SDSS low
metallicity galaxies referred in Kniazev et al. (2003), along with their reddening constants}
         \label{lines}

{
    $$  
         \begin{array}{cccccc}

            \hline
            \noalign{\smallskip}
            Object      &  [SIII] \lambda 6312 \AA & [SII] \lambda 6716 \AA 
& [SII] \lambda 6731 \AA & [SIII] \lambda 9069 \AA & C(H\beta) \\
                        
\hline
SDSS J0133+1342 & 0.9\pm0.2 & 9.9\pm0.2 & 8.0\pm0.4 & 5.8\pm 0.4 & 0.03 \\
KUG 0203-100    & --        & 23.0\pm1.1 & 18.6\pm1.1 & 19.6\pm 1.1 & 0.19 \\
HS 0822+3542    & 1.4\pm0.3 & 2.8\pm0.2 & 2.2\pm0.2 & 6.0\pm 0.2 & 0.17 \\	
I Zw 18 - NW    & 0.8\pm0.1 & 2.0\pm0.3 & 1.4\pm0.1 & 3.9\pm 0.3 & 0.05 \\	
I Zw 18 - SE    & 0.5\pm0.2 & 3.9\pm0.2 & 2.8\pm0.2 & 3.4\pm 0.5 & 0.12 \\	
SBS 1102+606    & 1.6\pm0.3 & 9.6\pm0.6 & 6.7\pm0.6 & 10.1\pm 0.6 & 0.15 \\		
A1116+517       & 1.3\pm0.3 & 7.9\pm0.2 & 5.3\pm0.2 & 7.7\pm 1.8 & 0.17 \\	
SDSS J1121+0324 & 1.3\pm0.4 & 12.2\pm0.8 & 9.6\pm0.8 & 4.2\pm 0.6 & 0.20 \\	  	
SDSS J1201+0211 & 1.0\pm0.2 & 5.4\pm0.2 & 3.7\pm0.2 & 5.0\pm 0.3 & 0.14 \\
CGCG 269-049    & 1.4\pm0.3 & 6.5\pm0.3 & 3.1\pm0.3 & 5.9\pm 0.7 & 0.10 \\

\noalign{\smallskip}
            
            \noalign{\smallskip}
            \hline
         \end{array}
$$
}
   \end{table*}

The strong nebular lines of sulphur are analogous to those of oxygen
and hence similar reasonings may be put forward regarding the use of
the S$_{23}$ parameter, S$_{23}$ = [SII]$\lambda\lambda$6717,6731 +
[SIII]$\lambda\lambda$9069,9532)/H$\beta$ (V\'{\i}lchez \& Esteban, 1996) as an alternative
abundance indicator (D\'{\i}az \& P\'erez-Montero, 2000), 
which in fact presents several advantages against
oxygen: 1) due to the longer wavelengths of the lines implied, its
relevance as a cooling agent starts at lower temperatures (higher
metallicities) what makes the relation to remain single-valued up to
solar abundances; 2) their lower dependence on electron temperature,
renders the lines observable even at over-solar abundances; 3) the
lines of both [SII] and [SIII] can be measured relative to nearby
hydrogen recombination lines, thus minimizing the effect of any
reddening and/or calibration uncertainties. On the negative side,
[SIII] lines shift out of the far red spectral region for redshifts
higher than 0.1. Given these properties, here we reccommend the use of
spectroscopy in the red-to-near infrared wavelength range in order to
derive physical properties and abundances, of HII regions, including
the lines from [SIII]$\lambda$ 6312 \AA , [SII]$\lambda$ 6717,31 \AA\
up to the [SIII]$\lambda$9069,9532 \AA . These lines can constitute a
convenient analogue in this wavelength range to the role played by the
[OII] and [OIII] lines in the optical. In additon, within this wavelength range 
the effects of the extinction are less severe, thus favouring the 
application of this calibration to the study of HII regions located 
in the Milky Way disk, in the inner regions 
of galaxies, or to star-forming galaxies/regions affected by a large amount of extinction.

In this paper, following earlier work by Christensen et al. (1997) and by Vermeij et al. (2002),  
we have explored the behaviour of the sulphur emission lines in order to provide a useful abundance 
calibration of S/H versus S$_{23}$ over the whole abundance range. This empirical calibration is based 
on the bright sulphur lines and encompasses the whole range of abundance currently found in HII region studies.
This calibration appears to be independent of both, O/H and O$_{23}$, and is especially suited for 
observations including (only) the red to near-infrared spectral range (from [SIII]$\lambda$ 6312 \AA , 
H$\alpha$ up to the 1$\mu$ CCD cut-off). For this wavelength regime, the sulphur lines can be easily scaled to 
their respective closest Balmer or Paschen recombination line, H${\lambda}$ --instead of to H${\beta}$-- 
and $S_{23}$ can be written as follows:

\[S_{23}  = S'_{23} x \left(\frac{H{\lambda}}{H{\beta}}\right)_o \]

\noindent where S'$_{23}$ is the scaled value of S$_{23}$ and
$(H\lambda/H\beta)_o$ is the theoretical case B recombination ratio.

The calibrations can be extended further to the infrared to include the [SIV] 10.52 $\mu$ line
to S$_{23}$ and defining a new parameter, S$_{234}$ (Oey \& Shields, 2000). 
However, the lack of [SIV] data renders difficult to check its reliability.

In the next section we present the sample of objects that we have used here to stablish the sulphur abundance 
calibration as well as all the selected objects for which reliable ISO [SIV] data have been published. 
Section 3 is devoted to the study of the ionisation correction factor (ICF),--not yet well stablished for sulphur--, 
using selected data of HII regions including [SIV] infrared line fluxes, together with predictions 
of photoionisation models. In Section 4 we present and discuss the proposed empirical calibration of 
S/H versus S$_{23}$ and we summarize our conclusions.


\section{The selected sample of objects and sulphur abundances}

Our sample is a combination of different emission line objects ionised by
young massive stars: diffuse HII regions in the Galaxy and the Local Group, Giant
Extragalactic HII regions (GEHR) and HII galaxies, and therefore does not include planetary
nebulae or objects with non-thermal activity. 
The emission line data along with their corresponding errors have been taken from the literature
for all objects except for the extremely metal-poor galaxies studied by Kniazev et al. (2003)
that present the strong [SIII] emission line at 9069 {\AA}.
The spectra of these objects have been taken from the data base of the DR3 of the Sloan Digital 
Sky Survey (SDSS\footnote{The SDSS Web site is http://www.sdss.org}) and 
the line fluxes have been measured and analyzed using the task SPLOT of the software package IRAF
\footnote{IRAF is distributed by the National Optical Astronomy Observatory}, following the
same procedure as in P\'erez-Montero \& D\'{\i}az (2003; hereinafter PMD03). We have found a good
agreement between the oxygen emission lines listed by Kniazev et al. (2003) and our measurements. 
The reddening corrected sulphur emission line fluxes, normalized to I(H$\beta$) = 100,
 which are relevant for this work are listed in table \ref{lines}, together with their corresponding
reddening constants and errors.

%
   \begin{table*}
      \caption[]{Sulphur abundances and values of log S$_{23}$ for the sample of objects.
Objects marked with $^a$($^b$) have S$^+$(S$^{2+}$) abundances derived using electron temperatures based on
models from t([OIII]). See text for details. }
         \label{Ab}

{
    $$  
         \begin{array}{cccc|cccc}

            \hline
            \noalign{\smallskip}
            Object      &  Ref.^a & 12+\log \left( \frac{S^++S^{2+}}{H^+} \right) & \log S_{23} &
	    Object      &  Ref.^a & 12+\log \left( \frac{S^++S^{2+}}{H^+} \right) & \log S_{23} \\
            \noalign{\smallskip}
            \hline

M51 - CCM10 & 1   & 6.95\pm0.10^a &  0.02\pm0.05  & N4A        & 4  & 6.74^a & 0.02    		     \\
M51 - CCM53 & 1   & 6.95\pm0.14^a &  0.01\pm0.06  & N79E       & 4  & 6.59^{a,b} & 0.00    		\\
M51 - CCM54 & 1   & 6.91\pm0.12^a &  0.08\pm0.05  & N191A      & 4  & 6.80^a & 0.14    		\\
M51 - CCM55 & 1   & 7.04\pm0.11^a &  0.05\pm0.06  & N80        & 4   & 6.02^{a,b} & -0.25    		\\
M51 - CCM57 & 1   & 6.82\pm0.15^a & -0.03\pm0.06  & N83        & 4   & 6.24^{a,b} & -0.09    		\\
M51 - CCM72 & 1   & 6.90\pm0.13^a & -0.03\pm0.06  & N13        & 4   & 6.20 & -0.08		\\	
NGC1232 - 04  & 2   & 6.43\pm0.21^a & -0.05\pm0.08  & N81        & 4   & 6.29^b & -0.07    \\		
NGC1232 - 05  & 2   & 7.01\pm0.16^a &  0.16\pm0.17  & N66        & 4   & 6.01^{a,b} & -0.31    		\\
NGC1232 - 07  & 2   & 7.02\pm0.21^a &  0.08\pm0.06  & NGC604 - A  & 5   & 6.82\pm0.03^{a,b}& 0.00\pm0.02   \\
NGC1232 - 14  & 2   & 6.84\pm0.16^a &  0.16\pm0.07  & NGC604 - C  & 5   & 6.78\pm0.02^{a,b} & 0.09\pm0.03   \\
NGC1365 - 05  & 2   & 6.92\pm0.08^a &  0.08\pm0.01  & NGC604 - D  & 5   & 7.03\pm0.13^a & 0.18\pm0.01   \\
NGC1365 - 08  & 2   & 6.81\pm0.08^a &  0.07\pm0.01  & NGC604 - E  & 5   & 7.00\pm0.14^a & 0.23\pm0.03   \\
NGC1365 - 14  & 2   & 6.96\pm0.11^a &  0.05\pm0.01  & NGC604      & 5   & 6.85\pm0.01^{a,b} & 0.06\pm0.01   \\
NGC1365 - 15  & 2   & 7.00\pm0.05^a &  0.23\pm0.01  & NGC604    & 8,5 & 6.79\pm0.09^b & -0.01\pm0.07   \\
NGC1365 - 16  & 2   & 6.69\pm0.13^a &  0.03\pm0.01  & NGC604    & 8,22 & 6.66\pm0.16 & -0.01\pm0.07   \\
NGC2997 - 04  & 2   & 6.85\pm0.18^a &  0.18\pm0.08  & NGC604    & 8,33 & 6.68\pm0.09^b & -0.01\pm0.07   \\
NGC2997 - 05  & 2   & 6.97\pm0.16^a &  0.18\pm0.08  & NGC595    & 8,33 & 6.78\pm0.15^b &  0.13\pm0.08   \\
NGC2997 - 06  & 2   & 6.92\pm0.16^a &  0.14\pm0.07  & IC 131     & 8,22 & 6.50\pm0.18^b &  0.06\pm0.08   \\
NGC2997 - 07  & 2   & 6.92\pm0.17^a &  0.13\pm0.08  & IC 131     & 8,33 & 6.65\pm0.15^b &  0.06\pm0.08 	\\
NGC2997 - 13  & 2   & 7.10\pm0.14^a &  0.15\pm0.07  & NGC588    & 8,22 & 6.42\pm0.14^b &  0.08\pm0.07 \\
NGC5236 - 03  & 2   & 6.92\pm0.17^a &  0.12\pm0.07  & NGC588    & 8,33 & 6.63\pm0.09^b &  0.08\pm0.07\\
NGC5236 - 06  & 2   & 7.00\pm0.20^a &  0.12\pm0.08  & NGC5471   & 8,24 & 6.18\pm0.13^b & -0.17\pm0.07\\
NGC5236 - 11  & 2   & 7.35\pm0.11^a &  0.30\pm0.07  & IC 10-2    & 8,23 & 6.61\pm0.06^b &  0.09\pm0.06\\
NGC5236 - 16  & 2   & 7.25\pm0.13^a &  0.25\pm0.07  & II Zw 40   & 8,7 & 6.13\pm0.05 & -0.22\pm0.06 \\
NGC628 - H13   & 3   & 6.41\pm0.07^a & -0.05\pm0.02 & II Zw 40   & 8,14 & 6.06\pm0.12 & -0.22\pm0.06\\
NGC1232 - CDT1 & 3   & 7.13\pm0.16^a & 0.11\pm0.03  & II Zw 40   & 8,21 & 6.11\pm0.06^b & -0.22\pm0.06\\
NGC1232 - CDT2 & 3   & 6.52\pm0.16^a & 0.21\pm0.03  & II Zw 40   & 8,23 & 6.14\pm0.06^b & -0.22\pm0.06\\
NGC1232 - CDT3 & 3   & 6.92\pm0.09^a & 0.16\pm0.02  & II Zw 40   & 8,24  & 6.12\pm0.06^b & -0.22\pm0.06 \\
NGC1232 - CDT4 & 3   & 6.93\pm0.07^a & 0.16\pm0.02  & NGC4861   & 8,6  & 6.37\pm0.12 & -0.02\pm0.12\\  
NGC2467         & 4   & 6.39^b & -0.03 		& NGC2363   & 8,26 & 5.83\pm0.14 & -0.40\pm0.13\\
\eta Car   & 4   & 6.99 & 0.19   		& I Zw 123   & 8,7 & 5.92\pm0.16^b & -0.27\pm0.15\\
M17        & 4   & 6.92^b & 0.20   		& Mrk 36     & 8,15  & 5.91\pm0.17^b & -0.19\pm0.17\\
M20        & 4   & 6.99^a & 0.20    		& Mrk 36     & 8,7 & 5.88\pm0.14^b & -0.19\pm0.17\\
NGC3576   & 4   & 7.14^b & 0.46    		& Mrk 600    & 8,24  & 6.00\pm0.14^b & -0.34\pm0.13\\
Orion 1    & 4   & 6.93 & 0.03    		& I Zw 18    & 8,23 & 5.49\pm0.15^{a,b} & -0.68\pm0.18\\ 
Orion 2    & 4   & 6.97^b & 0.26    		& I Zw 18    & 8,24  & 5.51\pm0.19^{a,b} & -0.68\pm0.18\\
N59A       & 4   & 6.60^a & -0.01    		& M101 - H681& 9 & 6.06\pm0.09^{a,b} & -0.20\pm0.07\\ 
N44B       & 4   & 6.84^a & 0.09    		& M51 - CCM10 & 10 & 7.16\pm0.20^a &  0.20\pm0.05 \\
N55A       & 4   & 6.88^a & 0.04    		& M51 -	CCM72 & 10  & 7.56\pm0.02^a &  0.41\pm0.00\\
N113D      & 4   & 6.90^a & 0.25    		& NGC2403 - VS35 & 11 &  6.77\pm0.17 & 0.10\pm0.03\\
N127A      & 4   & 6.83 & 0.12 			& NGC2403 - VS24 & 11 & 7.04\pm0.11 & 0.10\pm0.04\\	
N159A      & 4   & 6.53^{a,b} & 0.02    		& NGC2403 - VS38 & 11 & 6.87\pm0.09 & 0.06\pm0.02\\		
N214C      & 4   & 6.75^a & 0.12    		& NGC2403 - VS44 & 11 & 6.63\pm0.07 & 0.04\pm0.02\\

            \noalign{\smallskip}
            
            \noalign{\smallskip}
            \hline
         \end{array}
     $$ 
}
\begin{list}{}{}
\item[$^{\mathrm{a}}$] {\scriptsize References are: 
1. Bresolin et al., 2004; 2. Bresolin et al., 2005;
3. Castellanos et al., 2002; 
4. Dennefeld \& Stasinska, 1983; 5. D\'\i az et al., 1987; 6. Dinnerstein \& Shields, 1986; 
7. French, 1980; 8. Garnett, 1992; 9. Garnett \& Kennicutt, 1994; 10. Garnett et al., 2004; 11. Garnett et al., 1997;
12. Gonz\'alez-Delgado et al., 1995; 13. Gonz\'alez-Delgado et al., 1994
14. Guseva et al., 2000; 15. Izotov \& Thuan, 1988; 16. Izotov et al., 1994; 
17. Izotov et al., 1997; 18. Kennicutt et al., 2003; 19. Kinkel \& Rosa, 1994;
20. Kniazev et al., 2003; 21. Kunth \& Sargent, 1983; 22. Kwitter \& Aller, 1981;
23. Lequeux et al., 1979; 24. Pagel et al., 1992; 25. Pastoriza et al., 2003; 26. Peimbert et al., 1986; 
27. P\'erez-Montero \& D\'\i az, 2003;
28. Skillman \& Kennicutt, 1993; 
29. Skillman et al., 1994; 30. Terlevich et al., 1991; 31. Vermeij et al., 2002; 
32. V\'\i lchez \& Esteban, 1996; 33. V\'\i lchez et al., 1988; 34.This work.}
 
\end{list}
   \end{table*}
%
%
   \begin{table*}
    \setcounter{table}{1}
      \caption{Continued.}
         \label{Ab02}

{
    $$  
         \begin{array}{cccc|cccc}

            \hline
            \noalign{\smallskip}
            Object      &  Ref.^a & 12+\log \left( \frac{S^++S^{2+}}{H^+} \right) & \log S_{23} &
	    Object      &  Ref.^a & 12+\log \left( \frac{S^++S^{2+}}{H^+} \right) & \log S_{23} \\
            \noalign{\smallskip}
            \hline

NGC2403 - VS51 & 11 & 6.75\pm0.17 & 0.06\pm0.06 &  NGC3310 - M     & 25 & 6.85\pm0.01^b & 0.06\pm0.01   \\
NGC2403 - VS3  & 11 & 6.62\pm0.04 & 0.06\pm0.03 &  II Zw 40      & 27,7 & 6.06\pm0.04 & -0.28\pm0.02     \\
NGC2403 - VS49 & 11 & 6.59\pm0.17 & 0.06\pm0.05 &  II Zw 40      & 27,14 & 6.04\pm0.05 & -0.28\pm0.02    \\  
NGC2403 - VS48 & 11 & 6.60\pm0.26 & 0.08\pm0.09 &  II Zw 40      & 27,21 & 6.06\pm0.04 & -0.28\pm0.02  \\
NGC7714 - A    & 12 & 6.81\pm0.05^a & 0.19\pm0.01 & II Zw 40      & 27,23 & 6.10\pm0.04 & -0.28\pm0.02   \\
NGC7714 - N110 & 12 & 6.99\pm0.11^a & 0.18\pm0.05 & II Zw 40      & 27,24 & 6.06\pm0.04 & -0.28\pm0.02   \\
NGC7714 - B    & 12 & 6.31\pm0.06^a & 0.01\pm0.02 & Mrk 5         & 27,15 & 6.19\pm0.09 & 0.07\pm0.02\\
NGC7714 - C    & 12 & 6.41\pm0.18^a & 0.00\pm0.09 & SBS0749+568   & 27,17 & 6.04\pm0.11 & -0.17\pm0.04\\
NGC7714 - N    & 12 & 6.95\pm0.11^a & 0.22\pm0.05 & SBS0926+606   & 27,17 & 6.04\pm0.11 & -0.17\pm0.04\\
NGC2363 - A2   & 13 & 5.92\pm0.08 & -0.41\pm0.02 & Mrk 709       & 27,30 & 5.97\pm0.10 & -0.04\pm0.02\\  
M101 - H1013	& 18 & 7.12\pm0.07 & 0.21\pm0.02 & Mrk 22        & 27,16 & 6.12\pm0.11 & -0.11\pm0.05\\
M101 - H1105	& 18 & 6.95\pm0.06 & 0.18\pm0.02 & Mrk 1434      & 27,17 & 5.76\pm0.08 & -0.38\pm0.03\\
M101 - H1159	& 18 & 6.68\pm0.09 & 0.10\pm0.03 & Mrk 36        & 27,7  & 5.96\pm0.12 & -0.21\pm0.04\\   	
M101 - H1170	& 18 & 7.08\pm0.08 & 0.36\pm0.03 & Mrk 36        & 27,15 & 5.98\pm0.13 & -0.21\pm0.04\\   	
M101 - H1176	& 18 & 6.82\pm0.06 & 0.14\pm0.02 & VIIZw403      & 27,17 & 6.10\pm0.11 & -0.27\pm0.04\\  	
M101 - H1216	& 18 & 6.55\pm0.07 & 0.01\pm0.03 & UM461         & 27,15 & 5.74\pm0.06 & -0.35\pm0.02\\ 
M101 - H128	& 18 & 6.81\pm0.07 & 0.11\pm0.02 & UM462         & 27,15 & 6.13\pm0.10 & -0.13\pm0.04\\		 
M101 - H143	& 18 & 6.73\pm0.09 & 0.13\pm0.03 & Mrk209        & 27,17 & 5.94\pm0.10 & -0.28\pm0.04\\	
M101 - H149	& 18 & 6.72\pm0.07 & 0.12\pm0.02 & IZw18 - SE     & 28 & 5.31\pm0.12^b & -0.78\pm0.07\\ 	
M101 - H336 (S5)	& 18 & 7.10\pm0.09 & 0.21\pm0.03 & IZw18 - NW     & 28 & 5.13\pm0.10^b & -0.93\pm0.08\\	
M101 - H409	& 18 & 6.63\pm0.07 & 0.08\pm0.03 & UGC4483        & 29 & 5.77\pm0.13 & -0.39\pm0.05\\	
M101 - H67	& 18 & 6.60\pm0.14 & 0.07\pm0.03 &  N160 - A1      & 31 & 6.91\pm0.06 & 0.17\pm0.03\\ 	
M101 - NGC 5471A& 18 & 6.29\pm0.07 & -0.11\pm0.03&  N160 - A2     & 31 & 6.86\pm0.07 & 0.14\pm0.03\\
M101 - NGC 5471B& 18 & 6.31\pm0.08 & 0.02\pm0.03 & N159 - 5      & 31 & 7.01\pm0.10 & 0.22\pm0.03\\	
M101 - NGC 5471C& 18 & 6.42\pm0.08 & -0.05\pm0.03& N157 - B      & 31 & 6.87\pm0.09 & 0.24\pm0.04\\ 	
M101 - NGC 5471D& 18 & 6.51\pm0.10 & -0.02\pm0.03 & 30Dor - 1      & 31 & 6.72\pm0.08 & 0.06\pm0.02\\ 	
M101 - S5       & 19 & 7.31\pm0.04 & 0.31\pm0.00  & 30Dor - 2      & 31 & 6.80\pm0.09 & 0.11\pm0.02\\	
SDSS J0133+1342 & 20,34 & 5.58\pm0.13^a & -0.42\pm0.03 & 30Dor - 3      & 31 & 6.87\pm0.06 & 0.14\pm0.02\\	
KUG 0203-100    & 20,34 & 6.02\pm0.04^a & 0.04\pm0.05  & 30Dor - 4      & 31 & 6.79\pm0.08 & 0.12\pm0.02\\
HS 0822+3542    & 20,34 & 5.27\pm0.19^a & -0.59\pm0.01 & N11 - A        & 31 & 6.78\pm0.09 & 0.10\pm0.03\\	
I Zw 18 - NW    & 20,34 & 5.08\pm0.18^a & -0.78\pm0.04 & N83 - B        & 31 & 6.82\pm0.07 & 0.13\pm0.03\\	
I Zw 18 - SE    & 20,34 & 5.19\pm0.23^a & -0.73\pm0.06 & N79 - A        & 31 & 6.96\pm0.07 & 0.19\pm0.02 \\	
SBS 1102+606    & 20,34 & 5.67\pm0.19^a & -0.29\pm0.02 & N4 - A         & 31 & 6.79\pm0.07 & 0.10\pm0.02 \\		
A1116+517       & 20,34 & 5.49\pm0.31^a & -0.40\pm0.09 & N88 - A        & 31 & 5.92\pm0.10 & -0.32\pm0.03\\	
SDSS J1121+0324 & 20,34 & 5.49\pm0.31^a & -0.44\pm0.06 & N66            & 31 & 6.23\pm0.15 & -0.08\pm0.05\\	  	
SDSS J1201+0211 & 20,34 & 5.28\pm0.18^a & -0.58\pm0.02 & N81            & 31 & 6.22\pm0.10 & -0.17\pm0.03\\ 	
CGCG 269-049    & 20,34 & 5.39\pm0.21^a & -0.50\pm0.05 & S209       & 32   & 5.76\pm0.20^b & -0.35\pm0.07\\	
M101 - NGC5455    & 22   & 6.75\pm0.10^{a,b} &  0.14\pm0.05 & S127       & 32   & 6.42\pm0.29^b & -0.12\pm0.07 \\	
M101 - NGC5471    & 22   & 6.31\pm0.10^{a,b} & -0.06\pm0.05 & S128       & 32   & 6.40\pm0.29^b & -0.09\pm0.06 \\	
NGC3310 - A        & 25 & 6.82\pm0.03^a & 0.00\pm0.02   & MA2        & 33   & 6.94\pm0.34^b & 0.08\pm0.02  \\	
NGC3310 - B        & 25 & 6.78\pm0.02^b & 0.09\pm0.03   & NGC 604    & 33   & 6.76\pm0.03^b & 0.06\pm0.01  \\	
NGC3310 - C        & 25 & 7.03\pm0.13 & 0.18\pm0.01   & NGC 595    & 33   & 7.02\pm0.09^b & 0.36\pm0.02  \\	
NGC3310 - E        & 25 & 7.00\pm0.14 & 0.23\pm0.03   &  & & & \\

            \noalign{\smallskip}
            
            \noalign{\smallskip}
            \hline
         \end{array}
     $$ 
}
\begin{list}{}{}
\item[$^{\mathrm{a}}$] {\scriptsize  References are: 
1. Bresolin et al., 2004; 2. Bresolin et al., 2005;
3. Castellanos et al., 2002; 
4. Dennefeld \& Stasinska, 1983; 5. D\'\i az et al., 1987; 6. Dinnerstein \& Shields, 1986; 
7. French, 1980; 8. Garnett, 1992; 9. Garnett \& Kennicutt, 1994; 10. Garnett et al., 2004; 11. Garnett et al., 1997;
12. Gonz\'alez-Delgado et al., 1995; 13. Gonz\'alez-Delgado et al., 1994
14. Guseva et al., 2000; 15. Izotov \& Thuan, 1988; 16. Izotov et al., 1994; 
17. Izotov et al., 1997; 18. Kennicutt et al., 2003; 19. Kinkel \& Rosa, 1994;
20. Kniazev et al., 2003; 21. Kunth \& Sargent, 1983; 22. Kwitter \& Aller, 1981;
23. Lequeux et al., 1979; 24. Pagel et al., 1992; 25. Pastoriza et al., 2003; 26. Peimbert et al., 1986; 
27. P\'erez-Montero \& D\'\i az, 2003;
28. Skillman \& Kennicutt, 1993; 
29. Skillman et al., 1994; 30. Terlevich et al., 1991; 31. Vermeij et al., 2002; 
32. V\'\i lchez \& Esteban, 1996; 33. V\'\i lchez et al., 1988; 34. This work.}

\end{list}
   \end{table*}


For all the objects of the sample measurements of the emission lines of [SII] at 6717,31 {\AA}
and of [SIII] at 9069,9532 {\AA} exist, thus allowing the simultaneous determination of the  S$_{23}$
parameter and the abundances of S$^+$ and S$^{2+}$.

The physical conditions of the ionised gas in each sample object, including electron temperatures, 
electron density and sulphur abundances,
have been computed from the original emission line data using the same procedures as in 
PMD03, based on the five-level statistical
equilibrium model in the task TEMDEN and IONIC, respectively, of the software package IRAF 
(De Robertis, Dufour \& Hunt, 1987; Shaw \& Dufour, 1995). The atomic coefficients used are the
same as in PMD03 (see Table 4 of that work). Electron densities are determined
from the the [SII] $\lambda$ 6717{\AA} / $\lambda$ 6731{\AA} line ratio. Electron temperatures
have been calculated from the [SIII] ($\lambda$ 9069{\AA}+$\lambda$9532{\AA})/
$\lambda$ 6312{\AA} line ratio for all but 44 objects of the sample
for which the [OIII] ($\lambda$
4959{\AA}+$\lambda$5007{\AA})/$\lambda$ 4363{\AA} line ratio has been used. 
For these objects, marked with $^b$ in table 2, a theoretical relation between [OIII] and [SIII]
electron temperatures has been used:
\[t([SIII]) = 1.05t([OIII]) - 0.08\]

This relation is based on the grids of photo-ionisation models described in P\'erez-Montero \& D\'\i az (2005) 
and differs
slightly from the empirical relation found by Garnett (1992), mostly
due to the introduction of the new atomic coefficients
for S$^{2+}$ from Tayal \& Gupta (1999).  

Regarding [SII] temperatures, for those objects without the [SII] auroral lines
at 4068,4074 \AA. we have taken the approximation
t[SII] $\approx$ t[OII] as valid.
For 124 objects of the sample it has been possible to
derive t[OII] from the [OII]($\lambda$ 3726{\AA}+$\lambda$3729{\AA})
/$\lambda$ 7325{\AA} line ratio. \begin{footnote}{The [OII] $\lambda$7319{\AA}+$\lambda$7330{\AA} 
lines can have a contribution 
by direct recombination which increases with temperature. Using
the calculated [OIII] electron temperatures, we have estimated these contributions  to be less than 4 \% in all cases
and therefore we have not corrected for this effect}\end{footnote}. For
the rest of the objects of the sample, marked with $^a$ in table 2
, not presenting any auroral line in the low excitation zone,
we have resorted to the model
predicted relations between 
t([OII]) and t([OIII]) found in PMD03 that take explicitly into account the dependence of t([OII])
on electron density. This can affect the deduced abundances of $S^+/H^+$ 
by non-negligible factors, larger in all cases than the reported
errors. 

 For those objects for which multiple observations exist we have considered
each one of them as independent. 
The final quoted errors in the derived quantities have been calculated 
by propagating the measurement errors in the emission lines provided by
the different authors. This information is not provided for the objects from Dennefeld \& Stasi\'nska (1983;
reference 4). The ionic abundances of sulphur, S$^+$/H$^+$  and S$^{2+}$/H$^+$, together with
the values of the S$_{23}$ parameter for each object are given in table \ref{Ab}.

\begin{figure}[t]
   \centering
   \includegraphics[width=8.5cm,clip=]{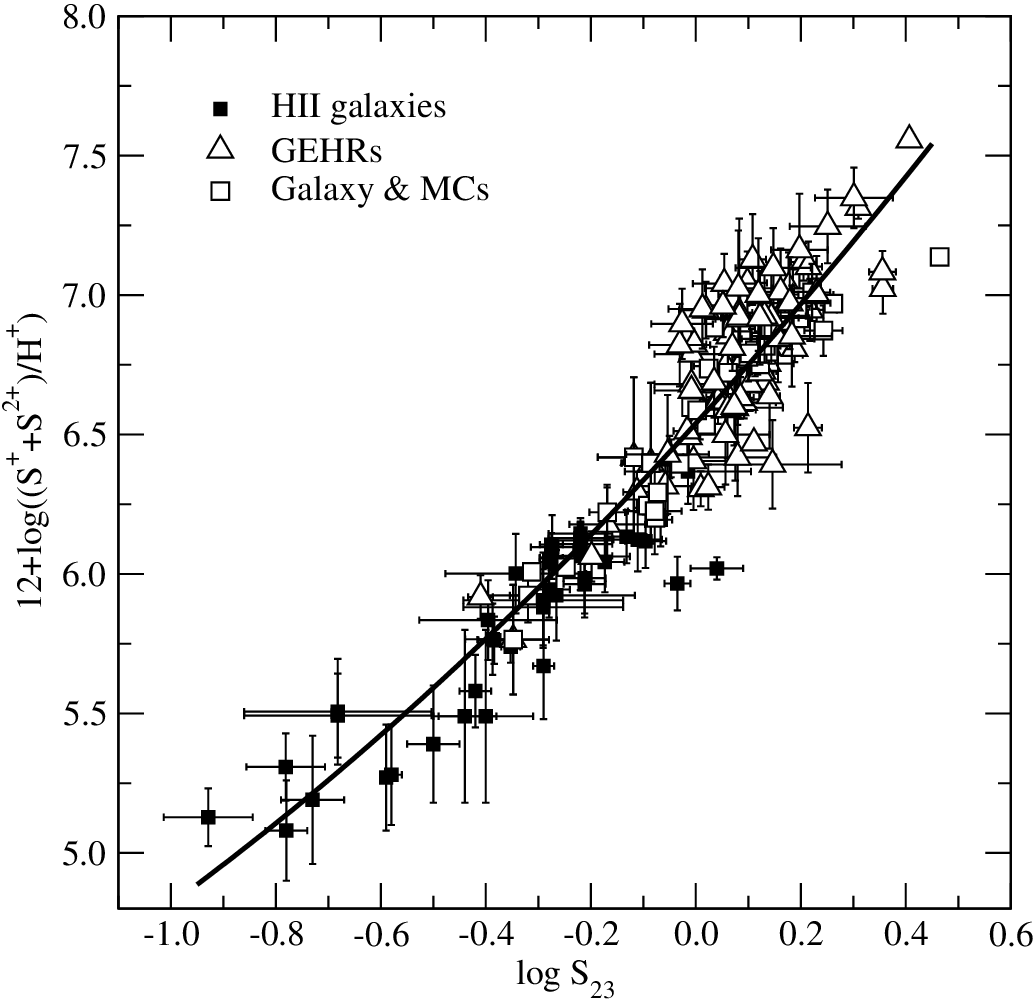}
   \caption{The calculated sulphur abundance for the objects of the sample
versus the S$_{23}$ parameter. The thick line represents the quadratic fit of the points.}
              \label{S23_S_01}
    \end{figure}

It is important to emphasize here that this database include objects covering all the 
abundance range from low metallicity HII galaxies, at 1/40 of the solar abundance, up 
to HII regions in the disks of spirals populating the high metallicity range, 
up to $\approx$ 3 Z$_\odot$. Although he data selected for this study have been obtained using different 
apertures and instrumental configurations and therefore do not constitute a homogeneous sample, they have been reanalysed and ionic abundances have been  derived in a homogeneous manner. Using this sample we can study on a firmer basis the 
empirical relationship between the sulphur abundance and the parameter S$_{23}$. In order 
to do that we explore first the sum of the abundances of S$^+$ and S$^{2+}$ as 
a function of S$_{23}$ and, on a second step, we will derive the calibration of 
the total abundance of sulphur as a function of S$_{23}$.  The result is plotted in 
figure \ref{S23_S_01}, showing a relationship with  very low scatter for which we have 
obtained the following quadratic fit:

\[12+\log\left(\frac{S^++S^{2+}}{H^+}\right) = 6.540 + 2.071 \log S_{23} + 0.348 \left( \log S_{23} \right) ^ 2 \]

\noindent with a typical dispersion of 0.17 dex, defined as the standard deviation of the residuals 
of the points.

After the advent of the ISO mission, it has also been possible to obtain data of
the [SIV] at 10.52 $\mu$ emission line for a selected sample of objects, thus allowing 
the derivation of the S$^{3+}$/H$^+$ ionic abundance ratio. We have collected data 
for 11 HII regions (plus one supernova remnant) in the Large and Small Magellanic Clouds 
(Vermeij et al., 2002) and for the HII galaxy Mrk209 (Nollenberg et al., 2002). 
For all these objects the $S^{3+}$ ionic abundances have been recalculated following 
the procedure described above and using the most recent atomic coefficients from 
Saraph \& Storey (1999). The calculated abundances are listed in table 3.

It has been a matter of debate for some time now what is the exact contribution of the 
S$^{3+}$/H$^+$ ionic abundance to the total abundance of sulphur. It is clear that an
efficient solution to this problem could be reached making use of the measurements of 
the infrared lines of [SIV] --not available for a large sample of objects yet--.

This question takes us straight into the issue of the ionisation correction factor, a
necessary step for the derivation of the empirical sulphur abundance calibration. In any 
case, it has been already pointed out that a one-zone ionisation scheme may provide some 
insight into the situation for not ionisation bounded HII regions of moderate to low ionisation, 
where the various ions coexist throughout the nebula (e.g. Pagel 1978). For this reason we 
have presented above the general correlation between S$^+$ + S$^{2+}$ and S$_{23}$. A  
study of the ionisation correction factor scheme for sulphur is presented in the next section.
  
%
   \begin{table}
      \caption{Abundances of S$^{3+}$ for the objects for which there are ISO observations of the
[SIV] 10.52 $\mu$ emission line and the corresponding ICF(S$^+$+S$^{2+}$).}
         \label{Ab2}
    $$  
         \begin{array}{cccc}

            \hline
            \noalign{\smallskip}
            Object      &  Ref.^a & 12+\log \left( \frac{S^{3+}}{H^+} \right) & ICF(S^++S^{2+})\\  
            \noalign{\smallskip}
            \hline
\\
Mrk209     & 1   & 5.93\pm0.20 & 1.97\pm0.62  \\
Mrk209^b   & 1 & 5.68\pm0.23  &  1.56\pm0.76 \\
N160-A1    & 2   & 6.08\pm0.11 & 1.15\pm0.05  \\
N160-A2    & 2   & 5.90\pm0.10 & 1.11\pm0.03  \\
N159-5     & 2   & 6.03\pm0.10 & 1.10\pm0.04  \\
N157-B     & 2   & 5.54\pm0.12 & 1.05\pm0.02  \\
30Dor-1    & 2   & 6.16\pm0.11 & 1.27\pm0.10  \\
30Dor-2    & 2   & 6.24\pm0.11 & 1.27\pm0.10  \\
30Dor-3    & 2   & 6.16\pm0.10 & 1.20\pm0.06  \\
30Dor-4    & 2   & 6.15\pm0.10 & 1.23\pm0.07  \\
N83-B       & 2  & 5.45\pm0.12 & 1.04\pm0.02  \\
N4-A        & 2  & 6.17\pm0.15 & 1.24\pm0.11  \\
N88-A       & 2  & 5.80\pm0.10 & 1.75\pm0.27  \\
N66         & 2  & 5.73\pm0.15 & 1.32\pm0.19  \\
N81         & 2  & 5.45\pm0.11 & 1.17\pm0.07  \\
\hline

\noalign{\smallskip}
            
            \noalign{\smallskip}
            \hline
         \end{array}
     $$ 
\begin{list}{}{}
\item[$^{\mathrm{a}}$] {\scriptsize The corresponding references for the emission line data
of [SIV] at 10.5 $\mu$ are: 1. Nollenberg et al., 2002; 2. Vermeij et al., 2002}
 \item[$^{\mathrm{b}}$] {\scriptsize Assumming a constant $S^{3+}/S^{2+}$ ratio from the [SIII] lines in the near-IR from PMD03}
\end{list}
   \end{table}

\section{The ICF Scheme}

Perhaps one of the most difficult aspects in the derivation of total
abundances is the question of the ionisation correction factor, ICF.
The ICF for sulphur accounts for the contribution of the ionic species
not detected in the optical.
In high excitation nebulae a large fraction of the sulphur can be found in the S$^{3+}$ stage, whose
prominent emission lines of [SIV] are observed in the mid-IR at 10.52 $\mu$. Therefore, in order to derive the total abundance of 
sulphur it is necessary to correct for the presence of S$^{3+}$ in those objects for which there are not  observations 
in the mid-IR, as follows.
 
\[\frac{N(S)}{N(H)} = ICF(S^++S^{2+}) \cdot \frac{N(S^++S^{2+})}{N(H^+)} =\]
\[= \frac{N(S^++S^{2+}+S^{3+})}{N(S^++S^{2+})} \cdot \frac{N(S^++S^{2+})}{N(H^+)}\]

The first proposed ICF scheme for sulphur (Peimbert \& Costero, 1969) was based on the similarity of the ionisation
potentials of O$^+$ (35.1 eV) and S$^{2+}$ (34.8 eV) 

\[ICF(S^++S^{2+}) = \frac{N(O)}{N(O^+)}\]

Nevertheless, some authors have pointed out that this relation has a
strong correlation with the ionisation degree of the nebula
(e.g. Barker 1978; Pagel 1978), thus implying an overestimation of the
sulphur abundance in nebulae with low electron temperature. Barker
(1980) proposed a new relation, based on the photoionisation models of
Stasi\`{n}ska (1978):

\[ICF(S^++S^{2+}) = \left[ 1-\left( 1-\frac{N(O^+)}{N(O)} \right)^\alpha\right]^{-1/\alpha} \]

\noindent for which he proposed a value of $\alpha$ = 3. The ICF from Peimbert \& Costero corresponds to a value for $\alpha$ = 1 in this expresion.
Later, Izotov et al. (1994), based on photoionisation models from Stasi\`{n}ska (1990) gave a fit for this ICF that,
in fact, is quite similar to the Barker formula for $\alpha$ = 2.

In figure 2 we show our computation of the ICF(S$^++S^{2+}$) as a
function of log ($O^+/O$) using the data listed in table 3, for three
different values of $\alpha$ =1,2,3. It is apparent in this plot that
most points are better matched for
$\alpha$ values between 2 and 3. Only one point, corresponding to a
HII galaxy (Mrk 209, solid square), shows an ICF for a value of $\alpha$ even lower than 2. Since the 
$S^{2+}$ abundances obtained from the [SIII] line at 18.71 $\mu$ from the ISO observations from
Nollenberg et al. (2002) are much higher than those obtained from the near-IR [SIII] (PMD03),
we have recalculated $S^{3+}$ abundances assuming a constant $S^{3+}$/$S^{2+}$ ratio.
This value is showed in table 3 and is represented as an open square in figures 2 and 3. 
Although showing a large error bar, 
the new value lies within the zone of $\alpha$ between 2 and 3, in better
 agreement to the values predicted by photo-ionisation models (PMD03).

   \begin{figure}[t]
   \centering
   \includegraphics[width=8.5cm,clip=]{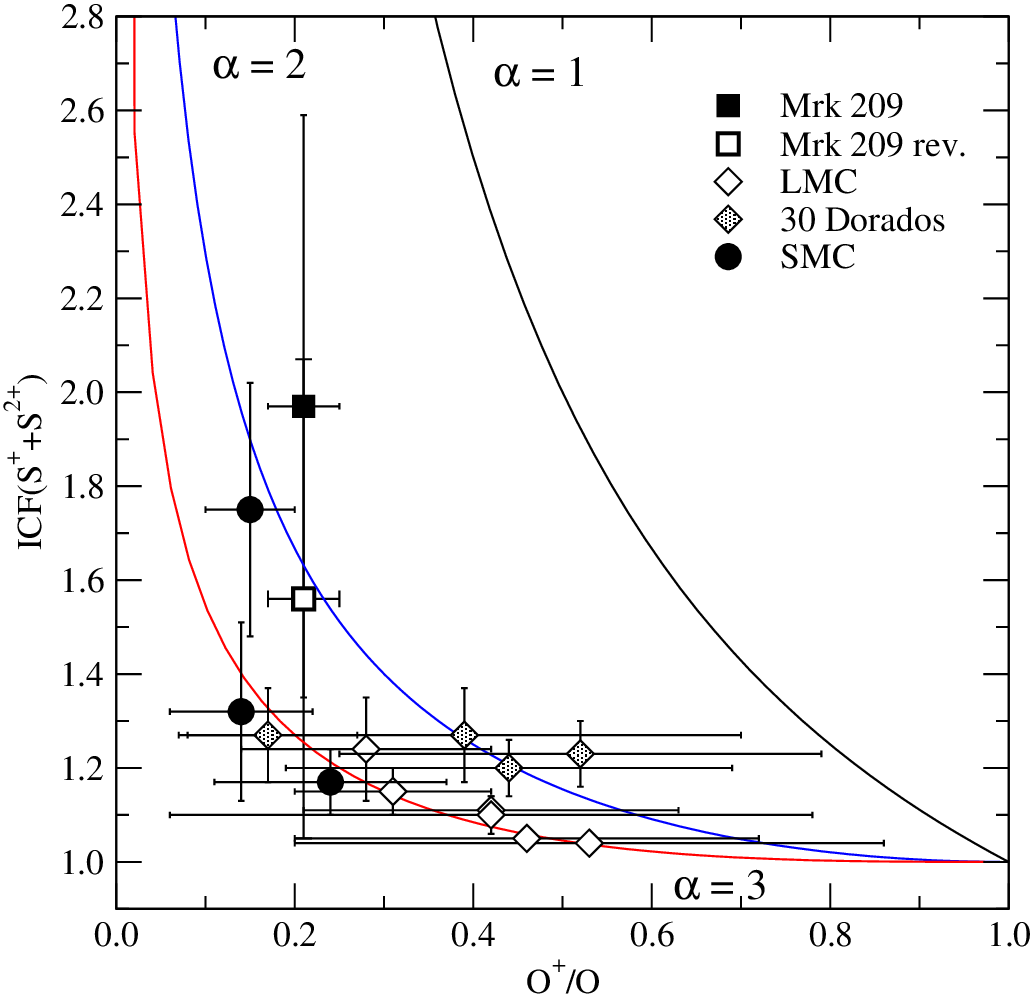}
   \caption{Ionisation correction factor for the sum of $S^{+}$ and $S^{2+}$ versus the $O^+/O$ ratio for the
sample of objects listed in table 3. The solid lines represent the Barker's formula for different values of the
$\alpha$ parameter. Squares represent different values for Mrk209 (see text for details), diamonds for
HII regions in the Large Magellanic Cloud and circles in the Small Magellanic Cloud.}
              \label{ICF_01}
    \end{figure}

\begin{figure}[t]
   \centering
   \includegraphics[width=8.5cm,clip=]{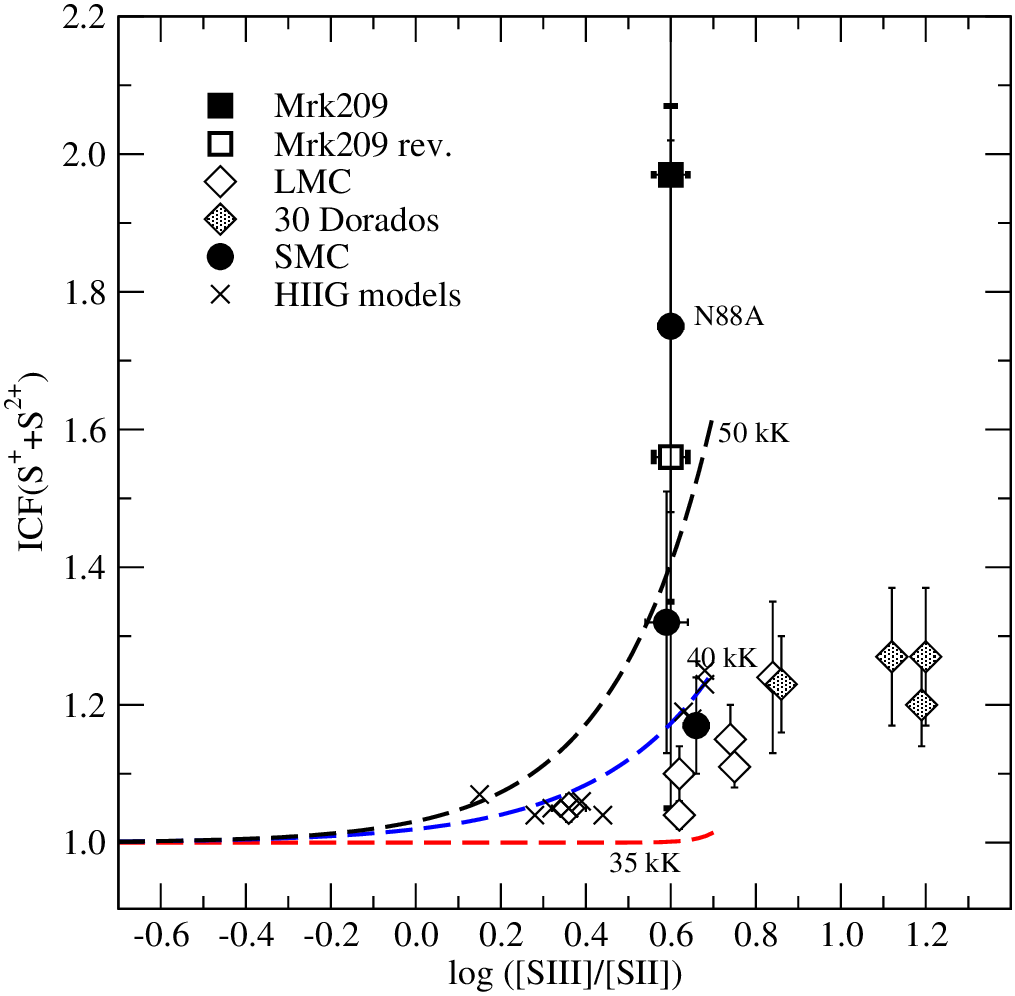}
   \caption{Ionisation correction factor for the sum of $S^{+}$ and $S^{2+}$ versus log ([SIII]/[SII]) for the
sample of objects listed in table 3. The dashed lines represent the results from photo-ionisation models for CoStar
single star atmospheres with the labelled effective temperature. Crosses represent results from photo-ionisation tailored models
of HII galaxies. The rest of symbols are the same as in figure 2.}
              \label{ICF02}
    \end{figure}

Since the relevant lines for sulphur abundance determination are in
the red to near infrared range, it is worth trying to relate the ICF
scheme of sulphur to the ionisation structure as seen in this
wavelength range. This aproach has the extra bonus of reducing the
effect of reddening in the line ratios currently used.  In order to
derive a new ICF scheme for sulphur based only on the red-to-near
infrared information, we need to combine available data and compute
new models for reference objects, for which we know all the relevant
ionic abundances of sulphur. We have computed this relation and is
shown in figure~\ref{ICF02}.  In this plot, we present the
ICF(S$^{+}$ + S$^{++}$) vs. log ([SIII]/[SII]) predicted by
photo-ionisation models using CoStar model atmospheres of 
different effective temperatures (35kK, 40kK, 50kK, P\'erez-Montero \& D\'\i az, 2005) 
and HII galaxy models (P\'erez-Montero \& D\'\i az, 2004),
together with all the observed points available.  Clearly, the ICF
predicted by a model with a single-star of Teff=35kK is always giving
ICF=1 no matter the excitation. The ICF predicted for objects
presenting log([SIII]/[SII]) $\le$0.2 remains small, 1.0$\le$ICF$\le$1.05. 
Above log([SIII]/[SII])=0.4, the ICF model predictions begin to
diverge. Up to some point, this behaviour seems to be followed by the
data. Giant HII regions points cluster around the
locus of 40kK models, except in the notable case of N88A, a SMC reddened, very young
HII region breaking out its natal cloud (Heydari-Malayeri et al., 1999).
In this plot this giant HII region is located closed to Mrk209, the only HII galaxy in table 3.
This fact indicates a hotter ionising source and a larger ICF. It is suggested here that
this behaviour could be consistent with the prediction from single
burst evolutionary models (e.g. Stasi\'nska et al. 2001) since typically
HII galaxies host younger ionizing clusters than giant HII regions (Terlevich et al. 2004).
Under this assumption, the equivalent
width of H$\alpha$, an age indicator, should provide a useful constraint.

\section{Discussion}

Earlier works by Christensen et al. (1997) and by Vermeij et al. (2002) have explored the abundance 
calibration of S/H versus S$_{23}$.
Christensen et al. (1997) proposed a linear S$_{23}$ calibration as follows 

\[12+\log\left(\frac{S}{H}\right) = 6.485 + 1.218 \log S_{23}\]

\noindent using data from giant HII regions and HII galaxies
available to this date, and complementary model predictions from
Stasi\'nska (1990). Though these points included some high metallicity
HII regions of M51 from Diaz et al. (1991), as well as lower
metallicity objects from Garnett (1992), still the whole abundance
range was not sufficiently well sampled. They claimed that more data were
needed before a definite conclusion could be drawn.

   \begin{figure}[t]
   \centering
   \includegraphics[width=8.5cm,clip=]{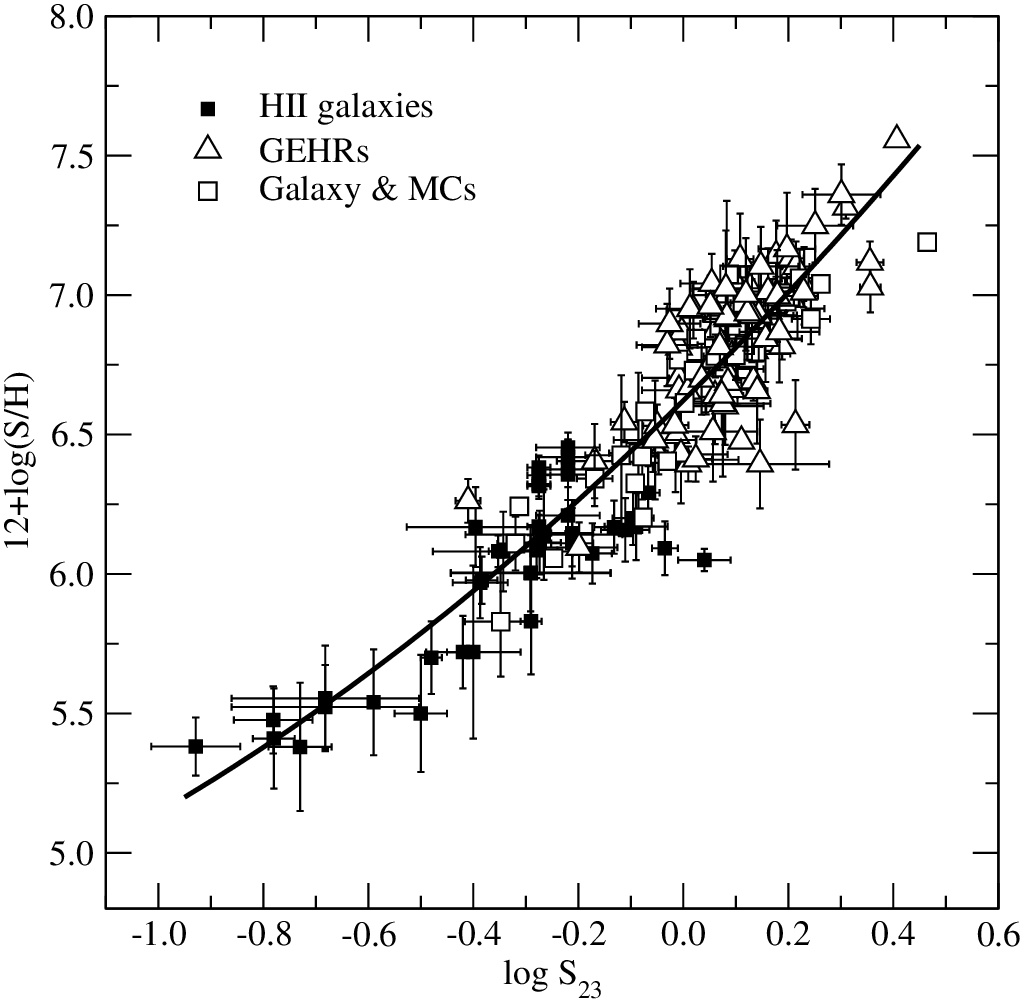}
   \caption{Calculated total sulphur abundance for the objects of the sample
versus the S$_{23}$ parameter. The thick line represents the empirical calibration proposed in this work}
              \label{s23S03}
    \end{figure}

Vermeij et al.(2002) derived sulphur abundances for a new data set of
optical and infrared spectra of HII regions in the Large and Small
Magellanic Clouds. This information allowed these authors to derive
all the ionic fractions of sulphur; including S$^{+3}$ from ISO
[SIV] 10.5 $\mu$ line for their sample objects. They present a
comparison of their S/H abundances points vs. S$_{23}$ and S$_{234}$
with model calculations for log U = 0, -1, -2, and -3. The following
relation was found between log S$_{23}$ and log(S/H)

\[\log\left(\frac{S}{H}\right) = -5.65 + 1.50 \log S_{234}\]

However no relation was proposed for S/H and S$_{23}$.

In this work,we present an empirical abundance calibration, based on
the bright sulphur lines, which encompasses the whole range of abundance
currently found in HII region studies. This calibration is firmly
based on an extended homogeneous data base.  In figure 4 we
present the relation for all the objects of our sample between the
S$_{23}$ parameter and the total abundance of sulphur, taking into
account the ionic abundances of S$^+$, S$^{2+}$ and the ICF
corresponding to the formula of Barker for $\alpha$ = 2.5 which is the
value that better fits the available points. The best quadratic fit to
the data gives:

\[12+\log\left(\frac{S}{H}\right) = 6.622 + 1.860 \log S_{23} + 0.382 \left( \log S_{23} \right) ^ 2 \]

\noindent with a dispersion of 0.185 dex in the range of -1.0 $\le$ log(S$_{23}$) $\le$ 0.5.

This calibration is, to first order, independent of both, O/H and O$_{23}$,
and is especially suited for observations including only the red to
near-infrared spectral ranges (from [SIII]$\lambda$ 6312 \AA ,
H$\alpha$ up to the 1$\mu$ CCD cut-off). Though the S$_{23}$ parameter is 
possibly double valued, like
R$_{23}$, the turnover region for S$_{23}$ is located above the range
of sulphur abundance currently found in disk HII regions. 

   \begin{figure}[t]
   \centering
   \includegraphics[width=8.5cm,clip=]{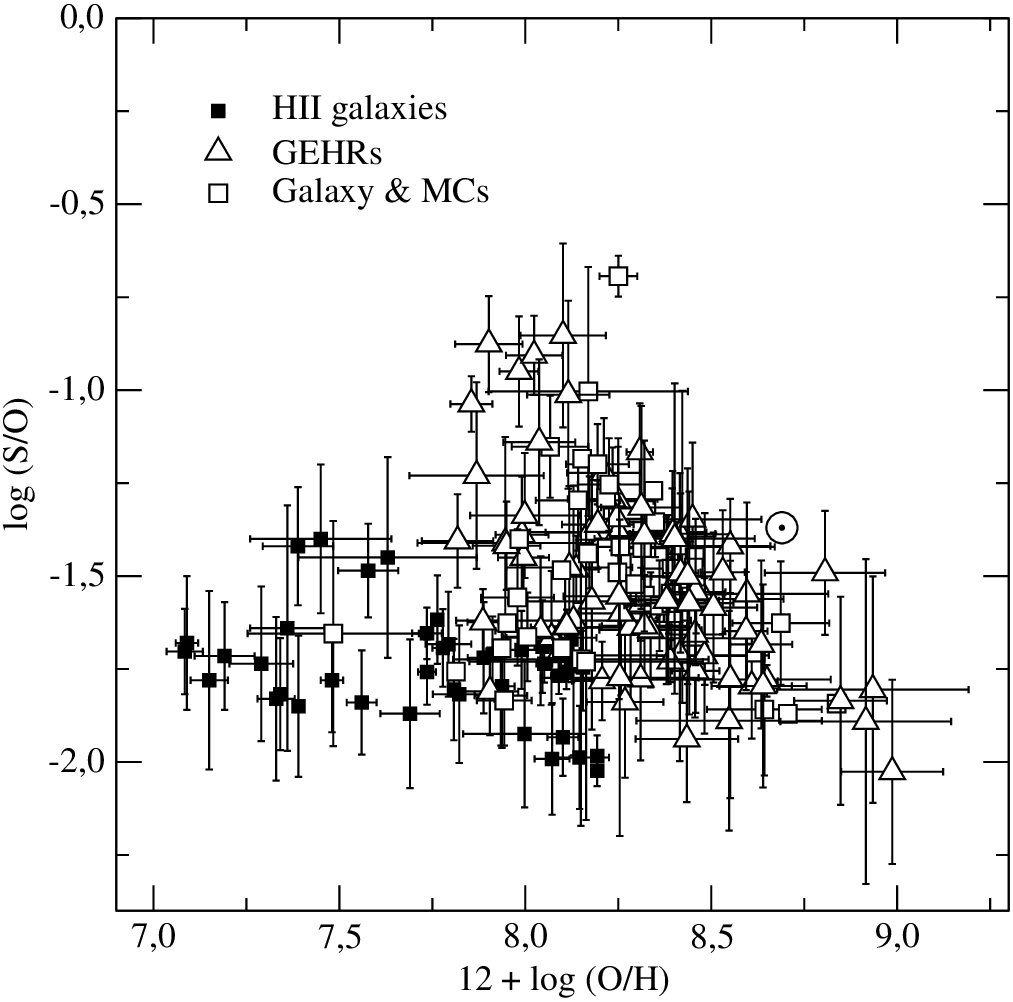}
   \caption{Representation of the quotient of S/O versus O/H for the compiled sample and the solar value }
              \label{SO_O}
    \end{figure}

The calibration is, to some extent, affected by the ICF calculation scheme.
However only 34\% of the objects in our sample show an ICF larger than 1.2
as derived from Barker's expression ($\alpha$ = 2.5) and all of them show
log ([SIII]/[SII]) $>$ 0.4, where model predictions are more uncertain. In fact,
the calibration of (S$^+$+S$^{2+}$)/H$^+$ vs. S$_{23}$ does not differ much
from that of S/H vs. S$_{23}$ (see figures 1 and 4) and the values obtained
from both for a given S$_{23}$ are within the quoted uncertainty in most cases.

This calibration makes possible the use of S/H as metallicity tracer in ionized
nebulae. Its translation however to an O/H abundance relies on the assumption that 
the S/O ratio remains constant at all abundances. This point though remains
controversial (see e.g.Lodders, 2003; Bresolin et al. 2004). Figure 5 shows
the S/O ratio vs. 12+log(O/H) for the objects in our sample. For all of them, the
O/H abundance has been calculated following the scheme presented in
P\'erez-Montero \& D\'{\i}az (2005). It can be seen from the figure that HII galaxy data
are consistent with a constant S/O ratio, but significantly lower than the solar ratio. Four
galaxies deviate from this trend (UGC4483, KUG 0203-100, Mrk709 and SDSS J1121+0324).
One of them, Mrk 709, also shows a large value of N/O (PMD03). 
Regarding disk HII regions the dispersion is much larger and the assumptions of a
constant S/O is highly questionable.

Summarizing, following earlier work by Christensen et al and Vermeij et
al, we have derived an empirical calibration of the abundance of S/H
as a function of the S$_{23}$ parameter, defined using bright sulphur
lines, which we recommend as a useful tool to derive S/H within a
large abundance range of 2 dex, keeping a statistical error of 0.18 
dex rms. This abundance range appears well suited to deal with objects
from low metallicity HII galaxies to high metallicity HII regions
located in the inner parts of the disks of spirals.

\begin{acknowledgements}
      This work has been partially supported by projects AYA-2004-08260-C03-02
and AYA-2004-08260-C03-03 of the Spanish National Plan for Astronomy and
Astrophysics.
\end{acknowledgements}

\end{document}